\documentclass[draft,grl]{agutex}

\usepackage{lineno}
\linenumbers*[1]

  \usepackage[dvips]{graphicx}
  \usepackage{ulem}

 \setkeys{Gin}{draft=false}
\authorrunninghead{Suweis et al.}

\titlerunninghead{{\sc Stochastic Soil Salinity}}

\begin{document}

\authoraddr{S. Suweis, Laboratory of Ecohydrology ECHO/IEE/ENAC/EPFL, \'Ecole Polytechnique F\'ed\'erale, Station 2, GR C1 515, Lausanne(CH)}
\authoraddr{A. Rinaldo, Laboratory of Ecohydrology ECHO/IEE/ENAC/EPFL, Lausanne (CH) and
Dipartimento IMAGE, Universit\`a di Padova, Padova (IT)}
\authoraddr{E. Daly, Department of Civil Engineering, Monash University, Clayton, VIC (AU) and National Centre for Groundwater Research and Training, Flinders University, Adelaide (AU)}

\authoraddr{S.E.A.T.M. Van der Zee, Soil Physics, Ecohydrology and Groundwater Management, Environmental Sciences Group, Wageningen University, Wageningen (NE)}
\authoraddr{A. Maritan, Dipartimento di Fisica Galileo Galilei, Universit\`a di Padova, Padova (IT)}
\authoraddr{A. Porporato, Department of Civil and Environmental Engineering, Duke University, Durham, NC (USA)}
%
%

\title{Stochastic Modeling of Soil Salinity}
%

%
%
\authors{S. Suweis, A. Rinaldo, S.E.A.T.M. Van der Zee, E. Daly, A. Maritan,\\ and A. Porporato}

%
%
%
%
%
%

%
%

\begin{abstract}
A minimalist stochastic model of primary soil
salinity is proposed, in which the rate of soil salinization is determined by the balance between dry and wet salt deposition and the intermittent leaching events caused by rainfall events. The long term probability density functions of salt mass and concentration are found by
reducing the coupled soil moisture and salt mass balance equation to
a single stochastic differential equation driven by multiplicative Poisson
noise. The novel analytical solutions provide insight on
the interplay of the main soil, plant and climate parameters responsible for long-term soil salinization.
In particular, they show the existence of two distinct regimes, one where the mean salt
mass remains nearly constant (or decreases) with increasing rainfall
frequency, and another where mean salt content increases markedly with increasing rainfall frequency.
As a result, relatively small reductions of rainfall in drier climates may
entail dramatic shifts in long-term soil salinization trends, with significant consequences
e.g. for climate change impacts on rain-fed agriculture.

\end{abstract}

%
%

\begin{article}

%
%

\section{Introduction}\label{I}

Large areas of cultivated land worldwide are affected by soil
salinity. \cite{Szabolcs1989} estimates that 10\% of arable land in
over 100 countries, and nine million km$^2$ are salt affected,
especially in arid and semi-arid regions
\citep{Tanji1989}. Salinity refers to large
concentrations of easily soluble salts present in water and soil on
a unit volume or weight basis (typically expressed as
electrical conductivity (EC) of the soil moisture in dS/m, i.e.
deciSiemens per meter at $25^{\circ}$ C; for $NaCl$ 1 mg/l $\sim $ $15\cdot10^{-4}$ dS/m). High salinity causes both
ion specific and osmotic stress effects, with important consequences
for plant production and quality. Normally, yields of most crops are not
significantly affected if EC ranges from $0$ to $2$ dS/m, while above
levels of $8$ dS/m most crops show severe yield reductions
\citep{Ayars1993,hillel2000}. Prevention or remediation of soil salinity is usually
done by leaching salts, and has resulted in the concept of leaching
requirement \citep{richards1954,Hillel1998,Schleiff2008}.
Alternative amelioration strategies by harvesting salt-accumulating
plants appear to be less effective \citep{Qadir2000}.

Salt accumulation in the root zone may be due to natural factors
(primary salinization) or due to irrigation (secondary
salinization). Several detailed numerical models have been developed to model soil
salinization \citep[e.g.][]{Eldin1987a,Tanji2006,UNSATCHEM}. Generally,
these models simulate unsaturated soil water flow via Richards
and solute transport equations. These models are more suitable for local and
short-term simulations, as they require precise soil
characterization and are computationally demanding. Moreover,
it is often difficult to identify cause-effect relationships or to
synthetically compare the effects of different parameter scenarios from their numerical simulations.

Vertically-averaged soil moisture and salt balance equations have also been used \citep{allison1994,hillel2000}. Despite their simplicity, these models have the advantage of parsimony, thus allowing a direct analysis of the interplay of the main processes, and provide an ideal starting point to include external, random hydroclimatic fluctuations in the analysis of long-term salinization trends. The goal of this Letter is to offer a first step in this direction.
With this purpose, here we present a minimalist model of soil primary salinization, describing
analytically the long-term dynamics of salt in soils caused by wet
(rain) and dry (aerosol) deposition. Our aim is to quantify the
salt mass and concentration probability density functions (pdfs) in
the root zone, and the probability of crossing the crops salt
tolerance threshold as a function of the main hydro-climatic
parameters. The model framework is potentially extendible to systems including salt input from groundwater and irrigation.

\section{Methods}\label{M}

Our starting point is a spatially lumped model \citep{bras1987}
for the vertically averaged dynamics of soil
moisture and salt in the root zone. As a first step we will not
consider input of salt due by irrigation or groundwater upflow.
Following \cite{iturbe1999a}, \cite{iturbe2004} and \cite{porporato2004}, rainfall ($R_t$)
is modeled as a marked point process with frequency $\lambda_{_P}$
and with daily rainfall depths exponentially distributed
with mean $1/\gamma_{_P}$. The averaged soil moisture dynamics
are modeled assuming constant (spatially and temporally averaged)
soil and ecohydrological parameters, i.e., root depth, $Z_r$, porosity,
$n$, and maximum evapotranspiration rate, $ET_{max}$. Assuming a
rain salt concentration $C_R$ and a constant input $\mathcal{M}_{d}$
of salt mass per unit ground area and per unit time by dry deposition, the
root-zone mass balance for soil moisture and salt mass $m$ is given
by:
\begin{linenomath}
\begin{eqnarray}
\label{soil_dyn} n Z_r \frac{ds}{dt} &=& -\,ET(s)\,-\,L(s)\, +\, R_t,\\
\label{Mass_balance} \frac{dm}{dt} &=& C_R R_t\,+\mathcal{M}_{d}\,-C L(s),
\end{eqnarray}
\end{linenomath}
where $C$ is the salt concentration in the root zone; $L(s)$ represents deep percolation, while
$ET(s)$ represents the losses resulting from plant transpiration and
soil evaporation. As in \citet{porporato2004}, $ET(s)$ is assumed to be linear in the
range of soil moisture comprised between the wilting point, $s_w$,
and a suitable soil moisture threshold $s_1$ (an effective
field-capacity threshold), at which $ET$ occurs
at the maximum rate $ET_{max}$. All the rainfall input that cannot be accommodated
is assumed to be lost as $L(s)$ at $s_1$. In this minimalist model the effect of salt-induced changes in osmotic potential may only indirectly be taken into account through an average reduction of $ET_{max}$. This is simply done here by keeping the same $ET_{max}$ for the two models (previous studies \citep{Viola2008} have shown that, in the absence of osmotic effects, the minimalist model should have artificially higher $ET_{max}$ to account for percolation losses below $s_1$). A full account of how reduction in evapotranspiration affects salinization patterns (reduced evapotranspiration in turn increases the available soil moisture and thus reduces the concentration of salt in the soil and increases leaching frequencies) will be given elsewhere.

\begin{figure}[h]
  \begin{center}
 \noindent\includegraphics[width=29pc]{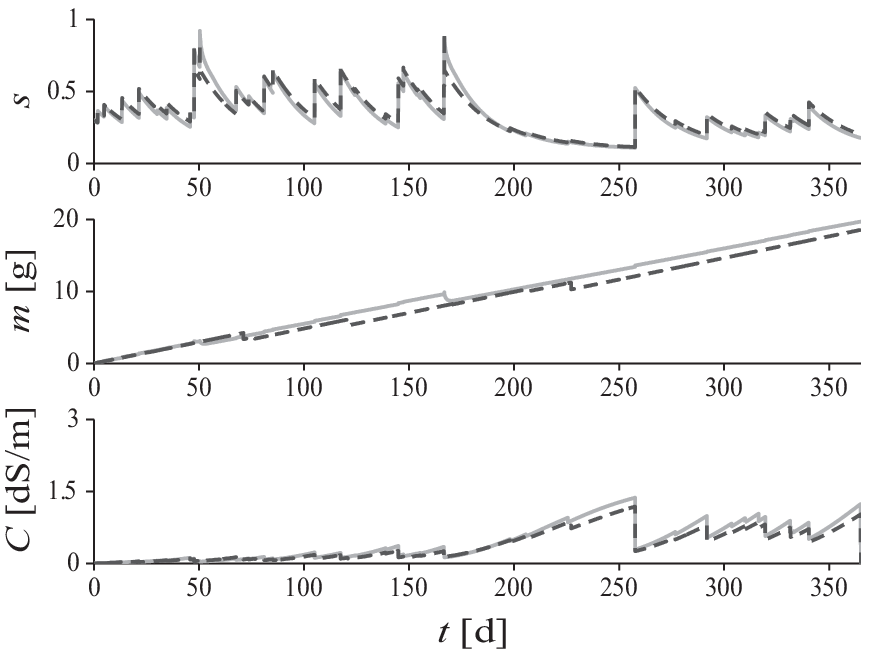}
 \caption{Comparison of soil moisture
and salinity models: (a) Temporal evolution (equation
(\ref{soil_dyn})) of $s(t)$, forced by intermittent rainfall
($\lambda_{_P} = 0.1$ d$^{-1}$  and $1/\gamma_{_P} =1.79$ cm). The blue
dashed line refers to the minimalist model, while the continuous
red line is the complete numerical model (see text for details).
(b) Temporal evolution of root-zone salt mass for the complete
numerical model (red line) and the minimalist model (blue dashed
line). (c) Temporal evolution of the corresponding specific salt
concentration $C(t)=m(t)/n Z_r s(t)$ in the root zone for the same
two cases of \ref{fig1}b. We transform the unit of measure of $C$ from mg/(cm m$^2$) to dS/m,
by using mg/(cm m$^2)=10^{-1}$ mg/l. The soil and vegetation
parameters employed for the simulation of the complete model are those typical for a sandy-loam soil, while the free parameters of the minimalist model are $s_1=0.8$, $b=0.6$. In particular for both models we used $n = 0.45$, $Z_r =
30$ cm, $s_w =0.1$, $ET_{max}=0.35$ cm/d, $C_R\approx 3$ mg l$^{-1}$ and $\mathcal{M}_d=54$ mg d$^{-1}$ m$^{-2}$ (coastal area).}\label{fig1}
  \end{center}
 \end{figure}

A complete numerical model, in which the impact
of osmotic stress in reducing $ET$ is explicitly included
\citep{bras1987}, has been also studied. Moreover, in the detailed model runoff takes place at saturation ($s=1$),
while percolation occurs for $s> s_{fc}$ (the soil moisture field capacity), and it is proportional to the soil hydraulic conductivity $K_{sat} s^{c}$, where $c$ is a soil-pore
connectivity index and $K_{sat}$ is the saturated hydraulic
conductivity \citep{iturbe2004}. A comparison between the results of the two soil
moisture models, presented in Figure \ref{fig1}a, suggests the viability
of the simplified model. Simulations for wetter climates confirm this result.

The system (\ref{soil_dyn}) and (\ref{Mass_balance}) can be further simplified if one
considers that the typical timescales for salt mass dynamics in the
root zone are orders of magnitude larger than the ones
characterizing rainfall (and thus wet deposition). Moreover, soil
moisture typically reaches steady-state conditions within a growing
season (e.g., $<5-7$ months), while the salt mass balance only does so
on much longer times scales (e.g., $>$ decades). Accordingly, at those
long timescales, say $T$, the salt mass input flux can be assumed to
take place at a constant rate, $\Upsilon$, that is $\int_t^{t+T}(\mathcal{M}_{d}+C_R R_t) dt' \sim
\mathcal{M}_{d} T+T\,C_R\lambda_{_P}/\gamma_{_P}=\Upsilon T$, and be interrupted by
instantaneous and unfrequent leaching events induced by percolation.
As a result, (\ref{Mass_balance}) can be rewritten as
\begin{linenomath}
\begin{equation}\label{Langevin}
   \frac{dm}{dt}=\Upsilon\,-\,\frac{m}{n\, Z_r\, s}\, L(s).
\end{equation}
\end{linenomath}
Leakage may then be modelled as a marked
point process, with percolation depths exponentially distributed with parameter $\gamma_{_P}$  \citep{botter2007a}. For reasons of analytical tractability, the percolation events are assumed to occur according to a Poisson process with frequency $\lambda$ given by the frequency of soil moisture crossing the
threshold $s=s_1$. This can be expressed in terms of the soil
moisture pdf as $\lambda=\rho(s_{1}) p(s_{1})$, where the term $\rho(s)=(ET(s)+L(s))/n Z_r$ represents the normalized
catchment-scale loss function (i.e. the total losses from the system due to evapotranspiration and leakage as a function of the soil moisture) \citep{iturbe2004}.
Adopting the soil moisture minimalist model (for which the pdf is a truncated gamma distribution, e.g., \cite{porporato2004}), the leaching frequency
is $\lambda= \eta\exp(-\gamma)
\gamma^{\lambda_{_P}/\eta}/\Gamma(\lambda_{_P}/\eta,
\gamma )$ \citep{botter2007a}, where $\Gamma(x,y)$ is the lower incomplete gamma function, $\eta=ET_{max}/(n Z_r
(s_1-s_w))$ and $\gamma=\gamma_{_P} n Z_r (s_1-s_w)$.

A leaching-efficiency parameter
$b$ is used to account for incomplete salt dissolution,
further assuming that the typical value of
soil moisture during leaching events can be approximated by the
value $s_1$. With the above assumptions, the dynamics of the
salt mass in the root zone can be described by a single equation
\begin{linenomath}
\begin{equation}\label{Langevin_mass}
   \frac{dm}{dt}=\Upsilon- m L'_t,
\end{equation}
\end{linenomath}
where $L'_t$ is a marked Poisson noise
\citep{VanDenBroeck1983} with frequency $\lambda$, and (dimensionless) exponential marks with mean
\begin{linenomath}
\begin{equation}\label{mu}
   \mu=\frac{b}{n\,Z_r\,s_{1}\,\gamma_{_P}}.
\end{equation}
\end{linenomath}
Figures \ref{fig1}b and \ref{fig1}c compare the results of both salinity models. The free parameters
$s_1$ and $b$ are fitted with respect to the complete model of salt mass and concentration, respectively.

From a mathematical viewpoint, equation (\ref{Langevin_mass}) is a
stochastic differential equation with multiplicative white (jump)
noise. In our case, since the soil solution can be considered in equilibrium during leaching events, one has to interpret
(\ref{Langevin_mass}) in the Stratonovich sense
\citep{VanDenBroeck1983}.
Accordingly, the normal rules of calculus are preserved, and equation (\ref{Langevin_mass})
can be transformed into
\begin{linenomath}
\begin{equation}\label{L_I}
\frac{dy}{dt} = \Upsilon e^{-y}-L'_t,
\end{equation}
\end{linenomath}
where $y(t)=\ln[m(t)]$.

\section{Results and Discussion}\label{RD}

The stationary solution of (\ref{L_I}) can be obtain as in \cite{iturbe1999a}. Then using the derived distribution for $m$, i.e., $p(m)=p(y)dy/dm$, we obtain the probability distribution for the salt mass in the root zone
\begin{linenomath}
\begin{equation}\label{p_mass_stat}
p(m)\,=\,\mathcal{N}\,\exp(-\frac{m\lambda}{\Upsilon})\, m^{1/\mu},
\end{equation}
\end{linenomath}
where
$\mathcal{N}=(\lambda/\Upsilon)^{\frac{1+\mu}{\mu}}/\Gamma(\frac{1+\mu}{\mu})$
and $\Gamma(x)$ is the Gamma function. Equation (\ref{p_mass_stat})
summarizes the soil salinity statistics as a function of climate,
soil and vegetation parameters.

\begin{figure}[h]
  \begin{center}
 \noindent\includegraphics[width=29pc]{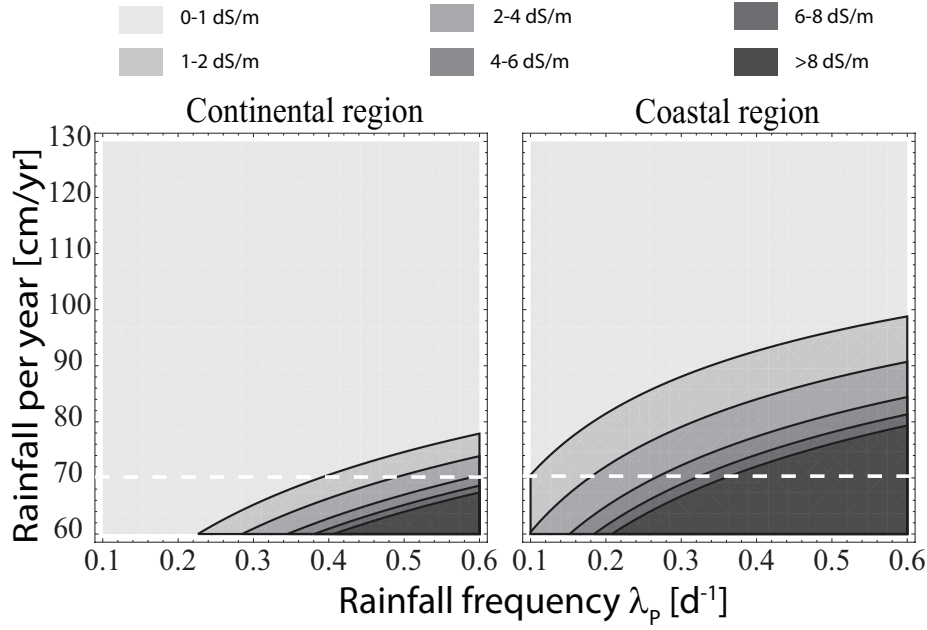}
 \caption{Contour plot of the asymptotic mean concentration of salt $\langle C \rangle$ from the exact solution of $\langle m \rangle$ as a function of yearly rainfall depth and frequency. The values reported in the legend refer to the corresponding salt concentration values with respect to the average soil moisture $\langle s\rangle$ (for its analytical expression see \cite{porporato2004}). The contour lines represent significant soil salinity values (1,2,...,8 dS/m). The parameter $\mu$ has been calculated through equation (\ref{mu}); the others are as in Figure \ref{fig1} for the coastal region, while for continental areas $\Upsilon\approx6$ mg d$^{-1}$ m$^{-2}$.}\label{fig2}
  \end{center}
 \end{figure}

Figure \ref{fig2} is a graphical
representation of the dependence of the mean salt concentration
    $\langle C\rangle=\langle m\rangle/n Z_r\langle s\rangle$
on the yearly rainfall and $\lambda_{_P}$. The contour-lines connect equal values of the
mean salt concentration in the soil, for a given input of salt $\Upsilon$.
The latter has been calculated for two different geographic regions. Typical salt inputs in coastal areas are $100-200$ kg/(ha yr) of salt, while values drop of an order of magnitude in continental regions \citep{hillel2000}.

Between the black region and the light gray ones in Figure \ref{fig2}a, the behavior of $\langle C \rangle$ changes substantially.
Above a certain total rainfall per year, the input of salt related to rainfall
frequency becomes immaterial as leaching effectively washes out the salt mass from
the root zone. For lower total rainfall values, however, the salt in the soil increases
with increasing $\lambda_{_P}$. For a given annual precipitation depth, with low rainfall frequencies,
rainfall events carry enough water to trigger leaching. Conversely, if $\lambda_{_P}$ is high,
evapotranspiration dominates, leaching is largely reduced, thereby causing salt accumulation in the root zone.
Therefore, $\langle m\rangle$ strongly increases with $\lambda_{_P}$.
Relatively small reductions of rainfall at the transition between these two regimes may entail a dramatic increase in long-term soil salinization.
Figure \ref{fig2} also shows the threshold of soil salinity below which vegetation is practically unaffected (e.g., $\langle C \rangle<2$ dS/m) and the thresholds above which regular (e.g. non-halophytic) vegetation is damaged (e.g., $\langle C \rangle>2$ dS/m). For coastal areas soil salinization may occur even in relatively more humid regions, especially when rainfall events are not very intense. On the contrary, in continental regions only arid climates may begin to develop soil salinization (in the absence of irrigation and groundwater input). Indeed, through our model one can evaluate the risk of soil salinization in rain-fed agriculture just by estimating the typical salt inputs, total rainfall per year and the rainfall frequency. For example, a rain-fed crop in a semi-arid climate (e.g., rainfall depth of $70$ cm/yr) in a continental region risks salinization only when rainfall events are not very intense (e.g., $\gamma_{_P}^{-1}\leq0.4$ cm or $\lambda_{_P}\geq0.48$ d$^{-1}$ ). If the same crop is located in a coastal area, salinization occurs for a wider range of rainfall parameters (e.g., $\gamma_{_P}^{-1}\leq1$ cm or $\lambda_{_P}\geq0.18$ d$^{-1}$ ).

\begin{figure}[h]
\begin{center}
\noindent\includegraphics[width=29pc]{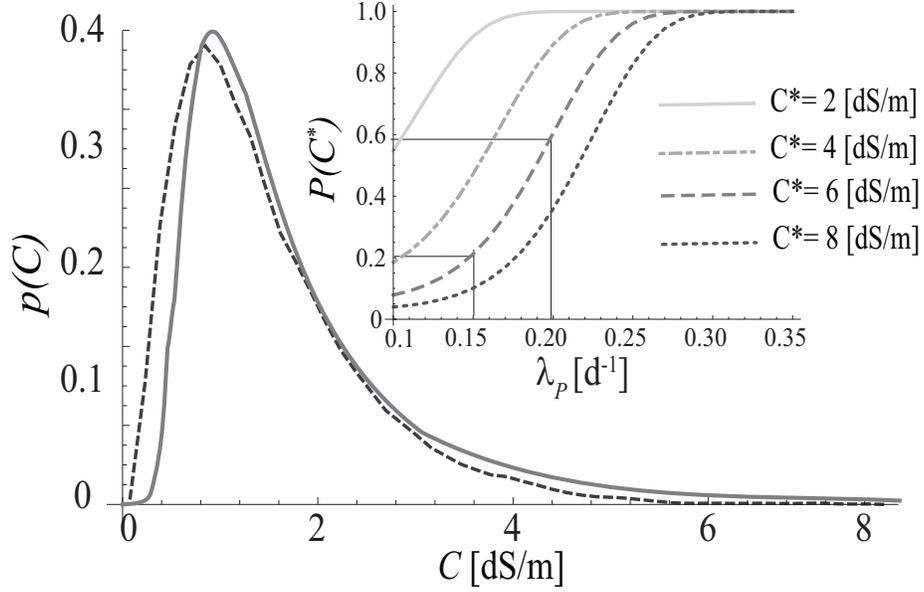}
\caption{Comparison
between the analytical form of $p(C)$ for the minimalist model obtained via equation (\ref{p(C)})
in which the free parameters ($b$ and $s_1$) have been fitted (solid line) and
the numerical simulations of the corresponding complete model (dashed line). The soil and hydro-climatic parameters are as in Figure \ref{fig1}. In the inset: probability of exceeding a soil salinization critical threshold $C^*$ ($\int_{C^*}^{+\infty}p(C)dC$) as a function of the rainfall frequency $\lambda_{_P}$.}\label{fig3}
\end{center}
\end{figure}

The solution (\ref{p_mass_stat}) may be used in conjunction
with soil moisture statistics to obtain a full characterization
of the salt concentration in the root zone.
Because one may safely assume that equations (\ref{soil_dyn}) and (\ref{Langevin})
are decoupled over short time scales, the soil moisture $s(t)$ and
the salt mass $m(t)$ may be treated as statistically
independent random variables. By observing that the salt
concentration in the root zone is equal to $C(t)=m(t)/n Z_r
s(t)$ and assuming $s_w\sim0$,
we find the stationary probability distribution of
the salt concentration $p(C)$ as the quotient distribution of two
independent random variables \citep{curtiss1941},
\begin{linenomath}
\begin{equation}\label{p(C)}
p(C)=\frac{\lambda(\frac{\Upsilon\gamma_{_P}}{C \lambda }+1)^{-1/\mu }(\frac{\Upsilon  \gamma_{_P}}{C \lambda +\Upsilon  \gamma_{_P}})^{\frac{\lambda_{_P}}{\eta}}(\Gamma(\frac{\lambda_{_P}}{\eta }+\frac{1}{\mu }+1)-\Gamma(\frac{\lambda_{_P}}{\eta }+\frac{1}{\mu }+1,n Z_r s_1(\frac{C \lambda }{\Upsilon }+\gamma_{_P})))}{\Gamma(1+\frac{1}{\mu }) (C \lambda +\gamma_{_P} \Upsilon )(\Gamma(\frac{\lambda_{_P}}{\eta})-\Gamma(\frac{\lambda_{_P}}{\eta },n Z_r s_1 \gamma_{_P}))}.\,
\end{equation}
\end{linenomath}
The comparison between analytical solutions
and numerical simulations (Figure \ref{fig3}) shows that the
analytical solution reproduces reasonably well the pdf of the complete model.

By integrating equation (\ref{p(C)}) from a given concentration value $C^*$ to infinity, one obtains the cumulative pdf of
$C$, $P(C^*)$, which is the
probability of having a salt concentration greater than a certain
critical concentration value, $C^*$, as a function of the
soil-plant-atmosphere parameters. The inset of Figure \ref{fig3} confirms the impact that
climate change may have on soil salinity.  Note, in particular, that such
an impact is marked only for semi-arid or drier climates
(see Figure \ref{fig2}). For example with a reduction
from $\lambda_{_P}=0.2$ to $\lambda_{_P}=0.15\,$d$^{-1}$, the probability of
crossing $C^*=6$ dS/m is more than tripled. When coupled to a crossing
analysis of concentration levels, the previous results may be used to evaluate the
risk of plant salt stress. The analytical form of the results makes
it suitable for computations of salinity risk at the global scale
as a function of few measurable parameters, and facilitates their coupling with
other models of long-term soil-plant biogeochemistry.

\section{Conclusions}

In this Letter we have presented an analytical approach to stochastic modeling
of soil salinity, where the complexity of the problem is reduced by employing
simplifying assumptions that permit us to describe high-dimensional, unpredictable components via suitable random terms.
By assuming time-averaged inputs of salt and instantaneous
percolation processes, a decoupling from soil moisture equation
results in a simplified stochastic mass balance equation for the
soil salt mass amenable to exact solution.

Soil salinity statistics are obtained as a function of
climate, soil and vegetation parameters. These can be
combined with soil moisture statistics to obtain a full
characterization of soil salt concentrations and the ensuing risk
of primary salinization.

This modeling framework can be extended to investigate additional salt inputs from irrigation and groundwater by modifying accordingly the average salt input parameter $\Upsilon$ and calculating the corresponding soil moisture pdfs (e.g. see \cite{vanderZee2008} for groundwater inputs and \cite{vico2009} for irrigation).


%

%

%
\begin{acknowledgments}
This research is supported by funds provided by the ERC Advanced
Grant RINEC-227612 and by SFN grant $200021\_124930/1$. AM acknowledges funds provided by Fondazione Cariparo (Padova, IT).
AP acknowledges the support of the Landolt \& Cie visiting
chair "Innovative Strategies for a Sustainable Future" at the
EPFL (CH) and the
collaboration grant with US Department of Agriculture,
Agricultural Research Service, Temple, TX; SvdZ acknowledges the
hospitality of ENAC/EPFL for his sabbatical stay.ED acknowledges the support of the Australian Research Council and the Australian National Water Commission.
\end{acknowledgments}


%
%
%
%
%

\end{article}


 \newpage

\newpage

 \newpage

\end{document}